\documentclass[]{elsart3p}
\usepackage{graphicx}
\usepackage{epsfig}
\usepackage{epstopdf}
\usepackage{amssymb}
\usepackage{subfigure}
\usepackage{setspace}

\renewcommand{\appendix}{
  \renewcommand{\section}{
      \newpage\thispagestyle{plain}
      \secdef\Appendix\sAppendix}
  \setcounter{section}{0}
  \renewcommand{\thesection}{\Alph{section}}
}

\newcommand{\Appendix}[2][?]{
  \refstepcounter{section}
  \addcontentsline{toc}{appendix}
      {\protect\numberline{\appendixname~\thesection} #1}
  {\flushleft\large\bfseries\appendixname\ \thesection\par

   \nohyphens\centering#2\par}
  \sectionmark{#1}\vspace{\baselineskip}
}

\newcommand{\sAppendix}[1]{
  {\flushleft\large\bfseries\appendixname\par
   \nohyphens\centering#1\par}
  \vspace{\baselineskip}
}

\newcommand{\nohyphens}{\hyphenpenalty=10000\exhyphenpenalty=10000\relax}

\begin{document}
\begin{frontmatter}

\title{Studies on fast triggering and high precision tracking with
  Resistive Plate Chambers}

\author[1]{G.~Aielli}
\author[2]{B.~Bilki}
\author[3]{R.~Ball}
\author[3]{J.W.~Chapman}
\author[1]{R.~Cardarelli}
\author[3]{T.~Dai}
\author[3]{E.~Diehl}
\author[3]{J.~Dubbert}
\author[3]{C.~Ferretti}
\author[3]{H.~Feng}
\author[2]{K.~Francis}
\author[3,4]{L.~Guan}
\author[4]{L.~Han}
\author[5]{S.~Hou}
\author[3]{D.~Levin}
\author[4,5]{B.~Li}
\author[3]{L.~Liu}
\author[1]{L.~Paolozzi}
\author[2]{J.~Repond}
\author[3]{J.~Roloff}
\author[1]{R.~Santonico}
\author[4]{H.Y.~Song}
\author[4]{X.L.~Wang}
\author[3,4]{Y.~Wu}
\author[2]{L.~Xia}
\author[3,4]{L.~Xu}
\author[6]{T.~Zhao}
\author[4]{Z.~Zhao}
\author[3]{B.~Zhou}
\author[3]{J.~Zhu}

\address[1]{University of Roma Tor Vergata and INFN Roma Tor Vergata, Roma, Italy}
\address[2]{Argonne National Laboratory, Argonne,USA}
\address[3]{University of Michigan, Ann Arbor, USA}
\address[4]{University of Science and Technology of China, Hefei, China}
\address[5]{Institute of Physics, Academia Sinica, Taipei, Taiwan}
\address[6]{University of Washington, Seattle, USA}

\date{October 24, 2012}

\begin{abstract}

We report on studies of fast triggering and high-precision tracking
using Resistive Plate Chambers (RPCs).  Two beam tests were carried
out with the 180 GeV muon beam at CERN using RPCs with gas gaps of
1.00 or 1.15 mm and equipped with readout strips with 1.27 mm pitch. 
This is the first beam test of RPCs with fine-pitch readout
strips that explores simultaneously precision tracking and triggering capabilities.  
RPC signals were acquired with precision
timing and charge integrating readout electronics at both ends of the
strips.  The time resolution was measured to be better than 600 ps and
the average spatial resolution was found to be 220 $\mu$m using charge
information and 287 $\mu$m using timing information.
The dual-ended readout allows the determination of the average and the
difference of the signal arrival times. 
The average time was found to be independent of the incident particle position
along the strip and is useful for triggering purposes. The time
difference yielded a determination of the hit position with a
precision of 7.5 mm along the strip. These results demonstrate the
feasibility using RPCs for fast and high-resolution triggering and tracking.

\end{abstract}

\begin{keyword}
RPC \sep trigger \sep tracking \sep time resolution \sep spatial resolution
\PACS 12.38.Qk \sep 12.15.Mm 
\end{keyword}
\end{frontmatter}

%%%%%%%%%%%%%%%%%%%%%%%%%%%%%%%%%%%%%%%%%%%%%%%%%%%%%%%%
\section{Introduction}
\label{sec-intro}

Resistive Plate Chambers (RPCs)~\cite{Santonico_1981} have been
rapidly adopted in large-scale particle physics experiments due to
their excellent timing capabilities, reliability, low cost, and
ability to easily scale to large areas.
In major collider and neutrino experiments, RPCs are mainly used as
trigger or time-of-flight devices with a typical time resolution of
$\mathcal{O}$(ns)~\cite{Aielli_2004,Abbrescia_2003,Arnaldi_2000,Williams_2002,Bertolin_2009,Giulio_2006,Zhang_2007}.
They are typically read out with centimeter-wide strips and have
spatial resolutions of $\mathcal{O}$(cm)~\cite{Cattani_2012,Arnaldi_2002}.
However, studies have shown that RPCs with narrow readout strips can
provide sub-millimeter spatial resolution using classic charge
interpolation \cite{Li_2012,Fonte_2012} or other techniques
~\cite{Cardarelli_2006}.  Such high resolution and fast RPCs could be useful as
trigger and precision tracking detectors in experiments at future lepton and hadron colliders.

To explore the feasibility of using RPCs for high-precision tracking
and fast trigger devices, two beam tests were carried out with the 180
GeV muon beam at CERN.  Single gas-gap RPCs equipped with fine-pitch
strips were used.  Both ends of the strips were read out using charge
Analog-to-Digital Converters (ADCs) and fast Time-to-Digital Converters (TDCs).  
A detailed description of the RPCs used in these beam
tests and their operations is described in
Sec.~\ref{sec-detector}.  The spatial resolution of muon hits in the
direction perpendicular to the readout strips (referred to as the
``primary coordinate''), was measured in the 2011 beam test and the
results are reported in Sec.~\ref{sec-primary-coord}.  The results
from the second beam test, focusing on measurements of RPC time
resolution, spatial resolution of muon hits in the direction along the
readout strips (referred to as the ``second coordinate"), and mean
signal arrival time from both ends of readout strips, are presented in
Sec.~\ref{sec-secondary-coord}. Conclusions are drawn in Sec.~\ref{sec-conclusion}.

%%%%%%%%%%%%%%%%%%%%%%%%%%%%%%%%%%%%%%%%%%%%%%%%%%%%%%%%
\section{RPC chamber description}
\label{sec-detector}

\begin{figure}[htbp]
  \begin{center}
    \includegraphics[width=0.95 \linewidth]{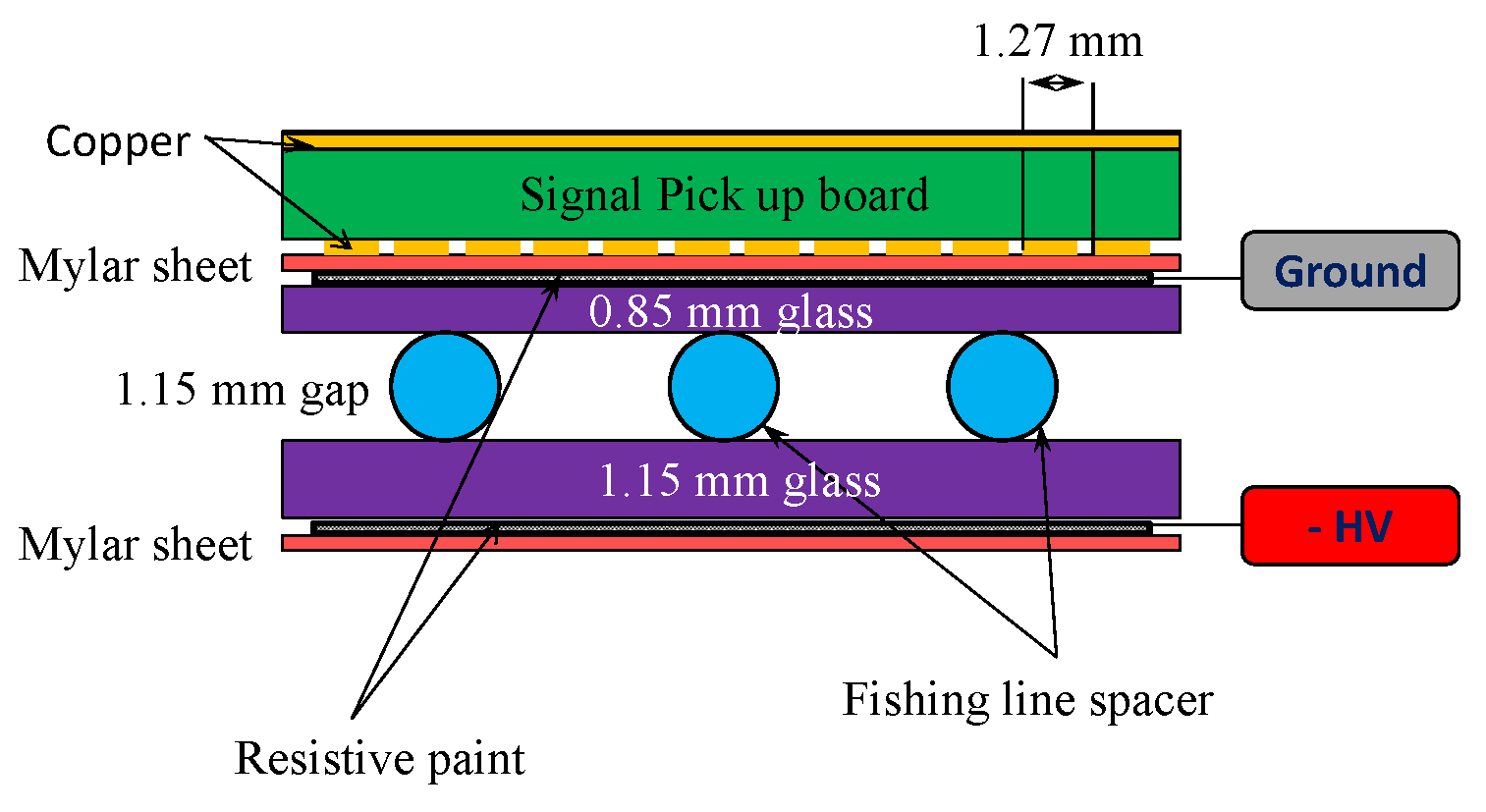}
  \end{center}
  \caption{Cross-sectional view of the glass RPC.}
  \label{chamber_scheme}
\end{figure}

%Single gas-gap RPCs with glass or Bakelite electrodes were used. 
Single gas-gap RPCs with glass or Bakelite electrodes were used in these tests. 
The glass chambers were 96 $\times$ 32 cm$^2$ and are depicted in
Fig. \ref{chamber_scheme}. Two glass plates were separated by 1.15 mm
using nylon monofilament placed along the longest dimension. The
ground-side glass plate was 0.85 mm thick and the side with the high
voltage applied was 1.15 mm thick. The outer surfaces of the glass
plates were painted with resistive paint having a surface resistivity
of 1-5 $M\Omega$/$\square$. The size of the Bakelite RPC was 20 
$\times$ 20 cm$^2$, with a structure similar to the glass
chambers but with a 1 mm gas gap and 2 mm thick electrodes.

Signal pickup boards with readout strips of 1.27 mm in pitch were
placed on the ground sides of the chambers, and signals were read out
via capacitive coupling. Negative high voltages were applied on the
opposite-side plates. A three-component gas blend of 
Freon(94.7\%):iC$_{4}$H$_{10}$(5\%):SF$_{6}$(0.3\%) was used.  The
typical operational voltages for glass and Bakelite chambers were
$6.5-7.0$ kV and $6.0-6.5$ kV respectively.  The dark currents were measured
to be less than 0.1 nA/cm$^2$ for the glass chambers and 0.6 nA/cm$^2$
for the Bakelite one.

%%%%%%%%%%%%%%%%%%%%%%%%%%%%%%%%%%%%%%%%%%%%%%%%%%%%%%%%
\section{Primary coordinate measurements}
\label{sec-primary-coord}
\subsection{Experimental setup and readout electronics}

\begin{figure}[htbp]
  \begin{center}
    \includegraphics[clip=true, width=0.95\linewidth]{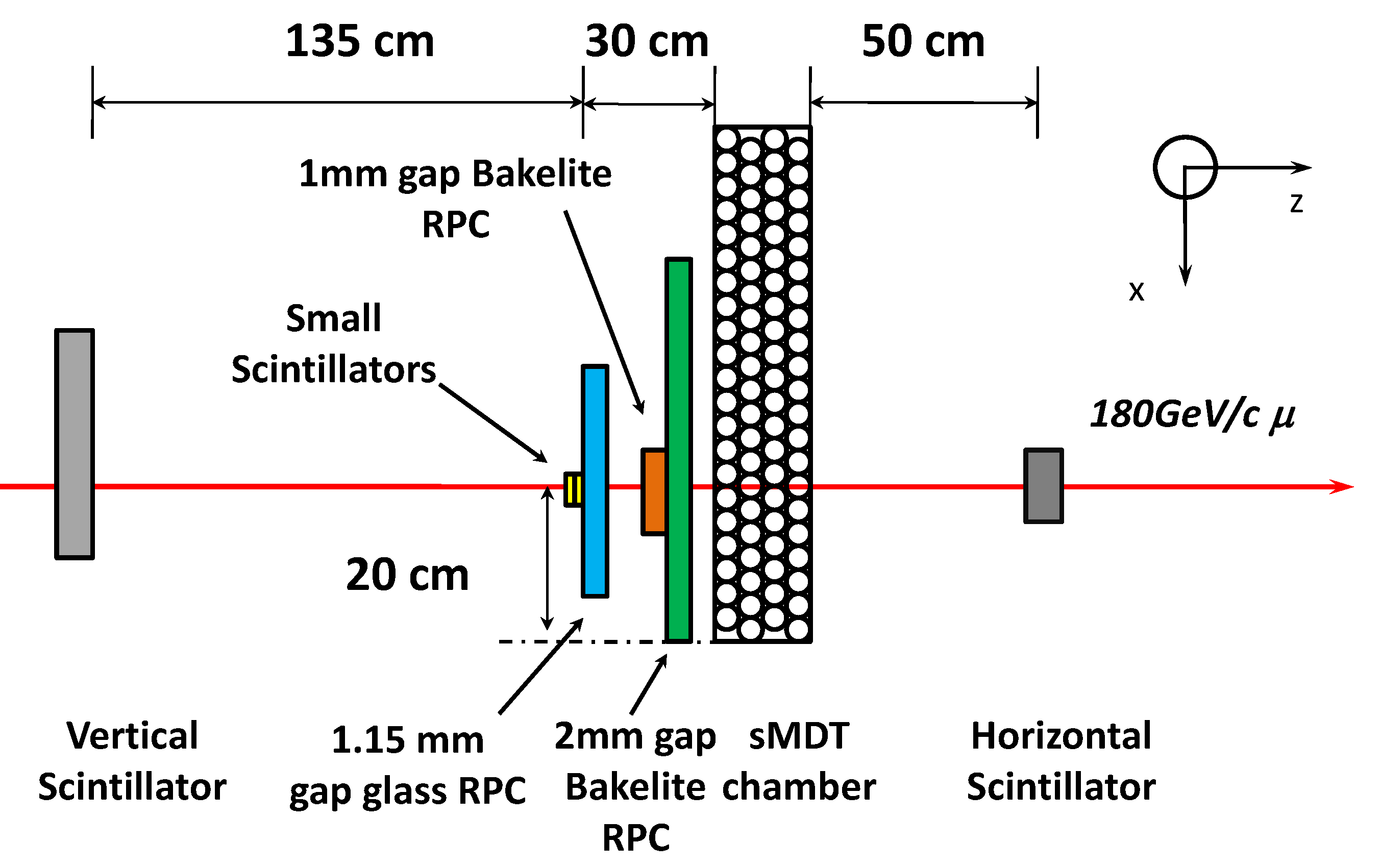}
  \end{center}
  \caption{Side view of the beam line set up for the primary coordinate measurement.}
  \label{2011tb_setup}
\end{figure}

The spatial resolution for the primary coordinate of the RPCs was
measured using 180 GeV muons from the SPS-H8 beam line at CERN in
October 2011. The schematic view of the experimental setup is
%illustrated 
shown in Fig.~\ref{2011tb_setup}. The glass RPC, accompanied
with two closely-spaced round scintillators, each 2 cm in diameter,
was placed upstream in the beam line. These two scintillators, smaller
than the readout area, were mainly used to measure the RPC
efficiency. Two Bakelite RPCs and a small-diameter monitored drift
tube (sMDT) chamber were installed at the downstream side. The sMDT
chamber~\cite{Bittner_2011} was made of eight layers of 15 mm diameter
drift tubes and was filled with an Ar(93\%):CO$_2$(7\%) gas mixture at
3 bar absolute pressure. The average spatial resolution of individual
drift tubes has been measured to be 120 $\mu$m. The sMDT chamber can
measure the direction of muon tracks with an angular resolution of 0.4
mrad~\cite{sMDT-perf} and thus provided precise
measurements of the incident muon tracks. Two additional large-area
scintillators were employed to give common trigger signals to all chambers.

In total, 72 strips from the glass RPC, terminated with 50 $\Omega$
resistors, were read out from both ends in three groups of 24
channels. Each group was connected to a low-noise ATLAS custom MDT
``mezzanine'' card containing three 8-channel
Amplifier-Shaper-Discriminator (ASD) chips \cite{Huth_1999} and a
24-channel TDC \cite{Arai_2002} chip which stored the arrival times of
signal leading and trailing edges in a large memory buffer. The pulse
height of the signal was measured by an ADC in the ASD chip and was
encoded as the time over threshold between leading and trailing edges
of the signal. The TDC has a least count of 0.78 ns. The data
from three mezzanine cards were formatted, stored in a large
derandomizing buffer and sent optically to the computer via a
local processor. The same system was used to read out the data from
the sMDT chamber using a common DAQ system.

\subsection{Cluster size and efficiency}

\begin{figure}[htbp]
  \begin{center}
    \includegraphics[width=0.9\linewidth]{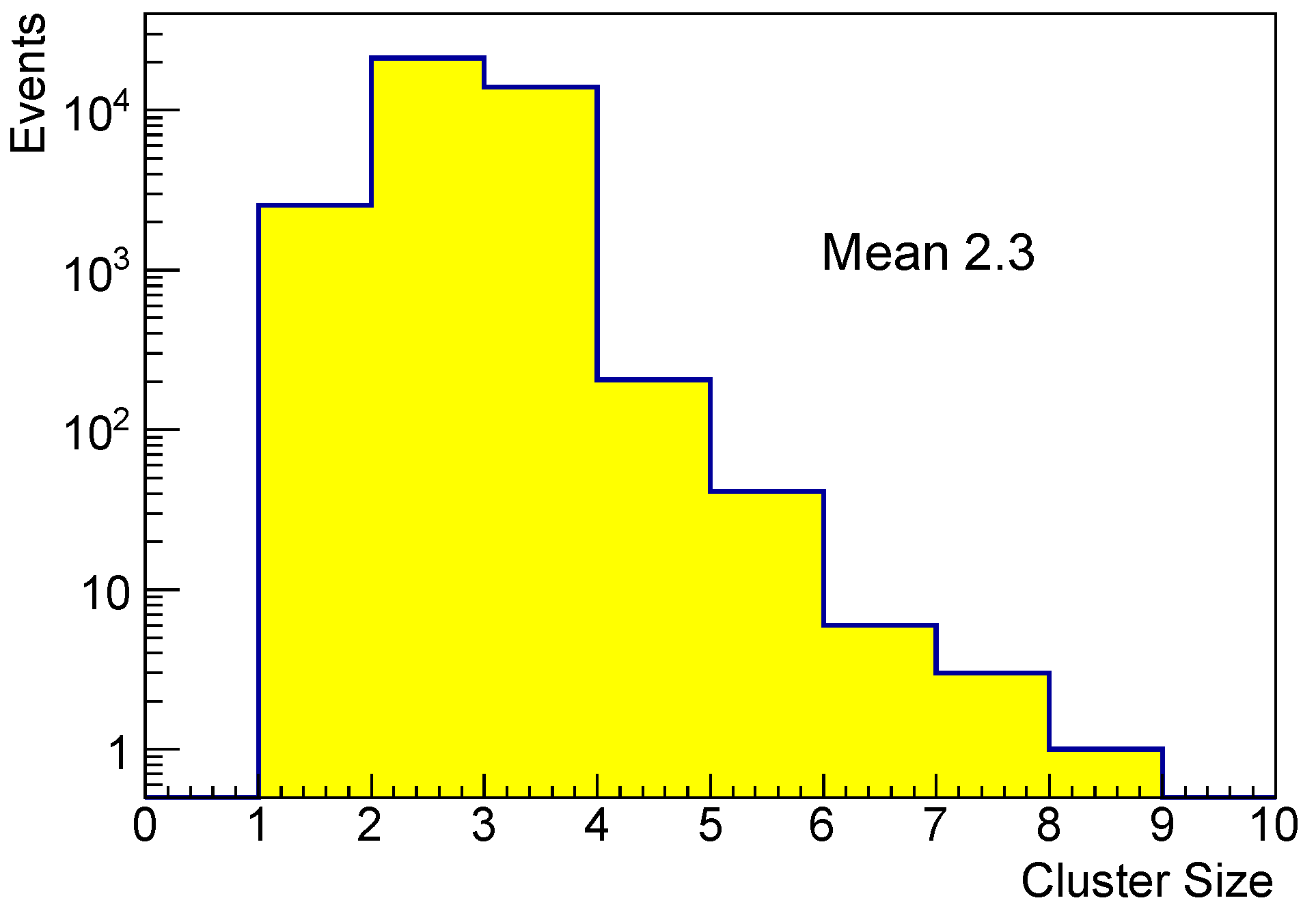}
  \end{center}
  \caption{Cluster size distribution with an operational high voltage
    of 6.5 kV.}
  \label{clustersize_6500v}
\end{figure}

\begin{figure}[htbp]
  \begin{center}
    \includegraphics[width=0.9\linewidth]{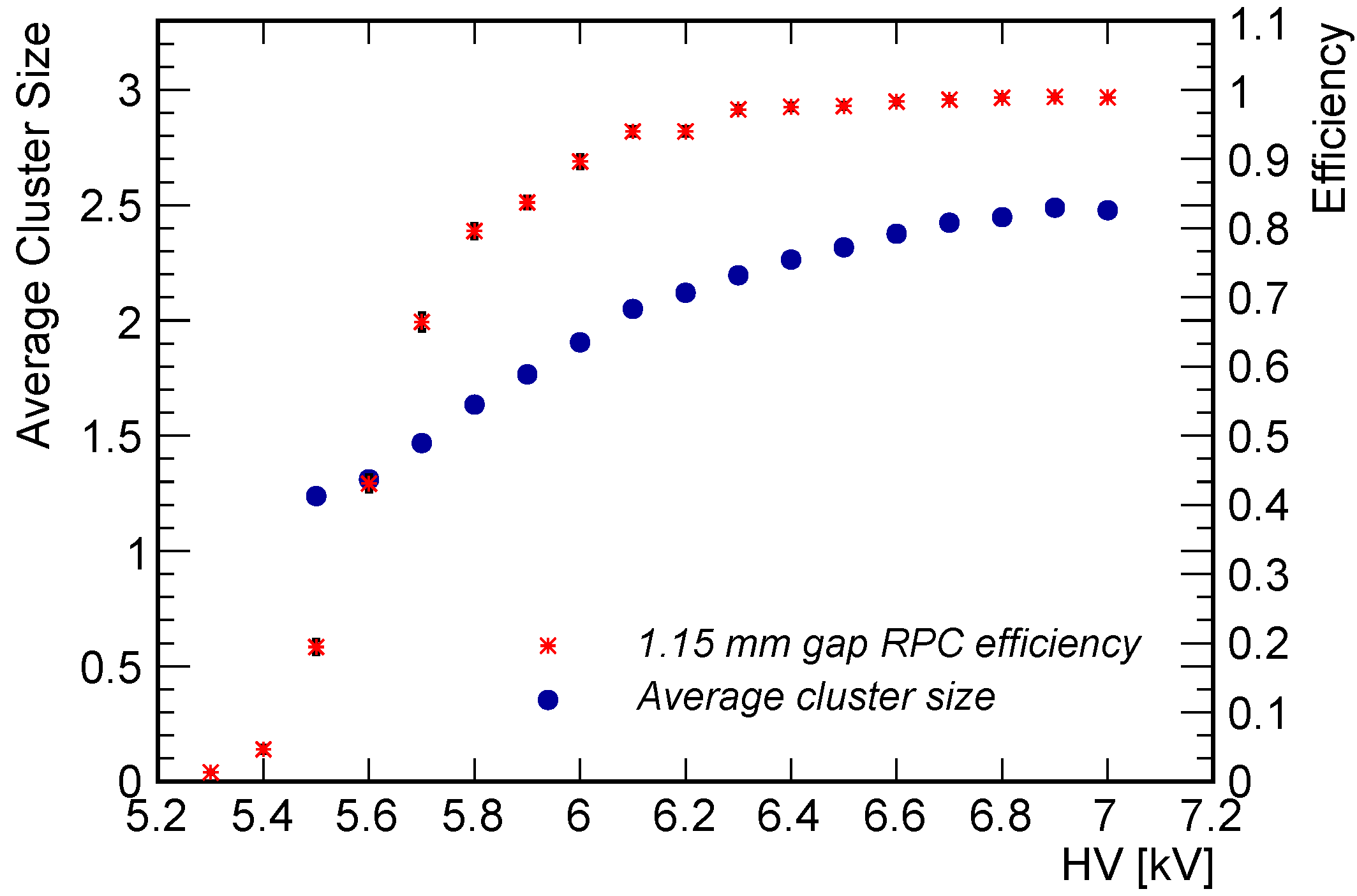}
  \end{center}
  \caption{The average cluster size and efficiency as a function of
    the applied high voltage for 1.15 mm gap glass RPC.}
  \label{cs_effi_hv}
\end{figure}

For events recorded by the glass RPC, the strip with the earliest
arrival time was selected, and a 20 ns time window was opened after
this earliest arrival time to look for hits on strips associated with the muon tracks.
A cluster is defined as a group of adjacent strips or combinations of
multiple strips separated by one missing strip. The distribution of
the cluster size with a high voltage of 6.5 kV is shown in
Fig.~\ref{clustersize_6500v}. The average cluster size was found to be
2.3 strips. Dependence of the average cluster size on the applied high
voltage is shown in Fig.~\ref{cs_effi_hv}. The cluster size increases
from 1.2 at 5.5 kV to 2.5 at 7.0 kV. 

The glass chamber efficiencies at different high voltages were
measured with a threshold of 5 fC on the total charge collected by
each strip, and the efficiency curve is also shown in Fig.~\ref{cs_effi_hv}. 

\subsection{Spatial resolution}

The glass RPC chamber was used for the primary coordinate
measurements.  Reference tracks from the sMDT chamber were
extrapolated to the RPC. The spatial resolution is then defined as the
difference between the measured position on the RPC and the expected
position from the reference track provided by the sMDT.

\begin{figure}[htbp]
  \begin{center}
    \includegraphics[width=0.9\linewidth]{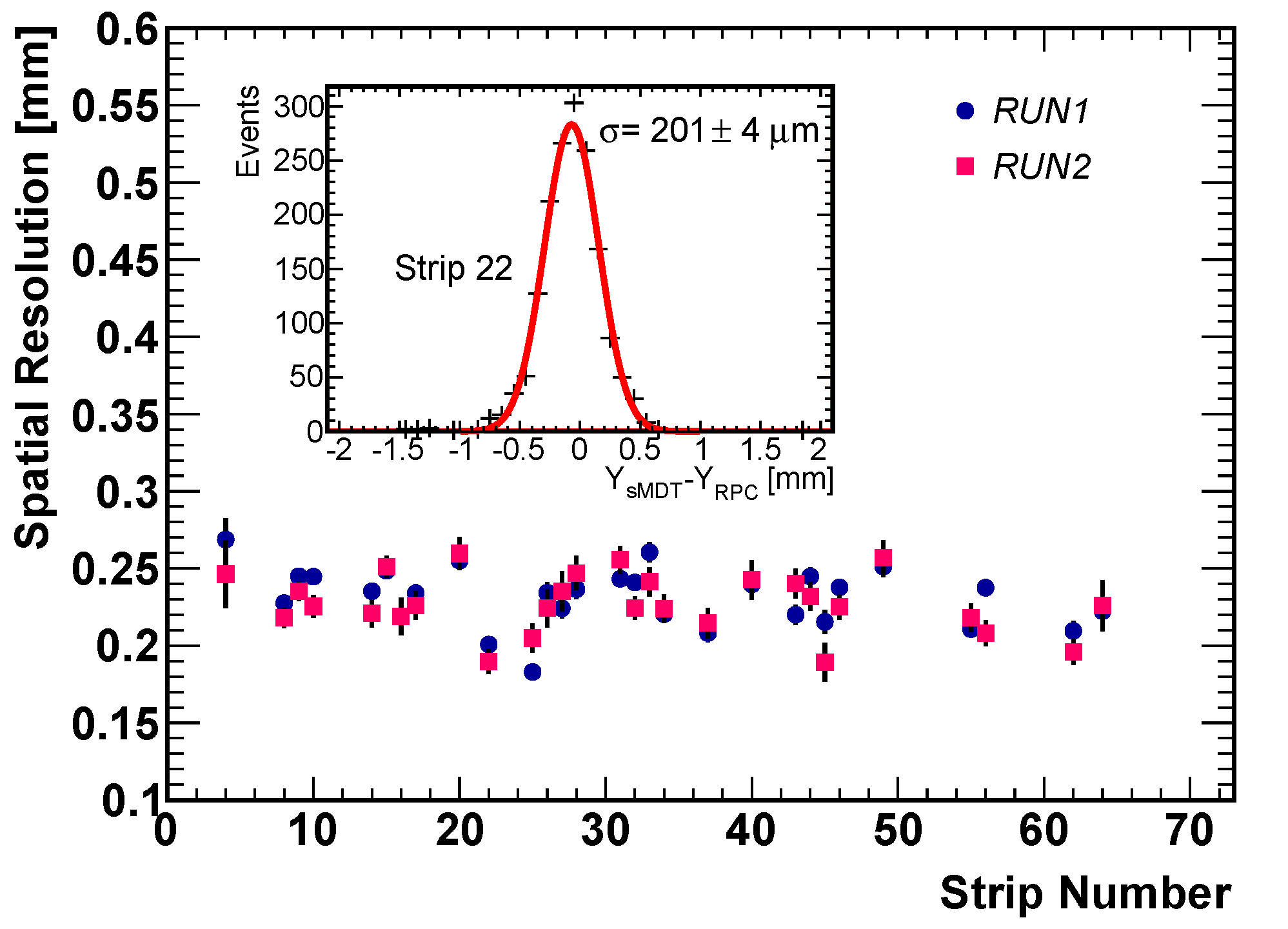}
  \end{center}
  \caption{RPC spatial resolution obtained using charge-weighted
    average positions with 1.27mm-pitch strips for two separate runs
    with same running conditions. Results have already subtracted 100
    $\mu$m track prediction uncertainty. The enclosed plot shows the
    residual distribution for a typical strip.}
  \label{offline_space_res}
\end{figure}

The track hit positions in the RPC were reconstructed with two
different methods. For the first approach, the hit positions were
determined as the weighted average of the strip center positions using
strip charge ADCs as weights. Since a dedicated sMDT calibration was
not available during the beam test and the RPC was placed almost half
a meter away from the sMDT center, a conservative assumption of 100
$\mu$m uncertainty for the predicted track hit positions was made. 
The RPC spatial resolutions after subtracting this uncertainty,
are shown in Fig.~\ref{offline_space_res} as a function of the hit
position in the readout area for two separate runs with same running conditions. 
Some strips were missing due to the geometric cuts of the
sMDT reference tracks. The residual distribution for a typical strip
is also shown as an enclosed plot in Fig.~\ref{offline_space_res}.
The average spatial resolution for all strips was found to be 220 $\mu$m.  
However, it should be noted that the resolution obtained here is not the
ultimate resolution that can be achieved with this detector.  The ADC
value measured with the ATLAS MDT electronics used in this beam test
has a logarithmic dependence on the total charge deposited.  Hence,
the spatial resolution of the centroid calculated using these ADC
values is worse than the one that would be reconstructed using the
actual accumulated charge.  Moreover, the RPC signals begin to
saturate the MDT readout electronics and thus degrade the calculated
hit position precision.

\begin{figure}[htbp]
  \begin{center}
    \includegraphics[width=0.9\linewidth]{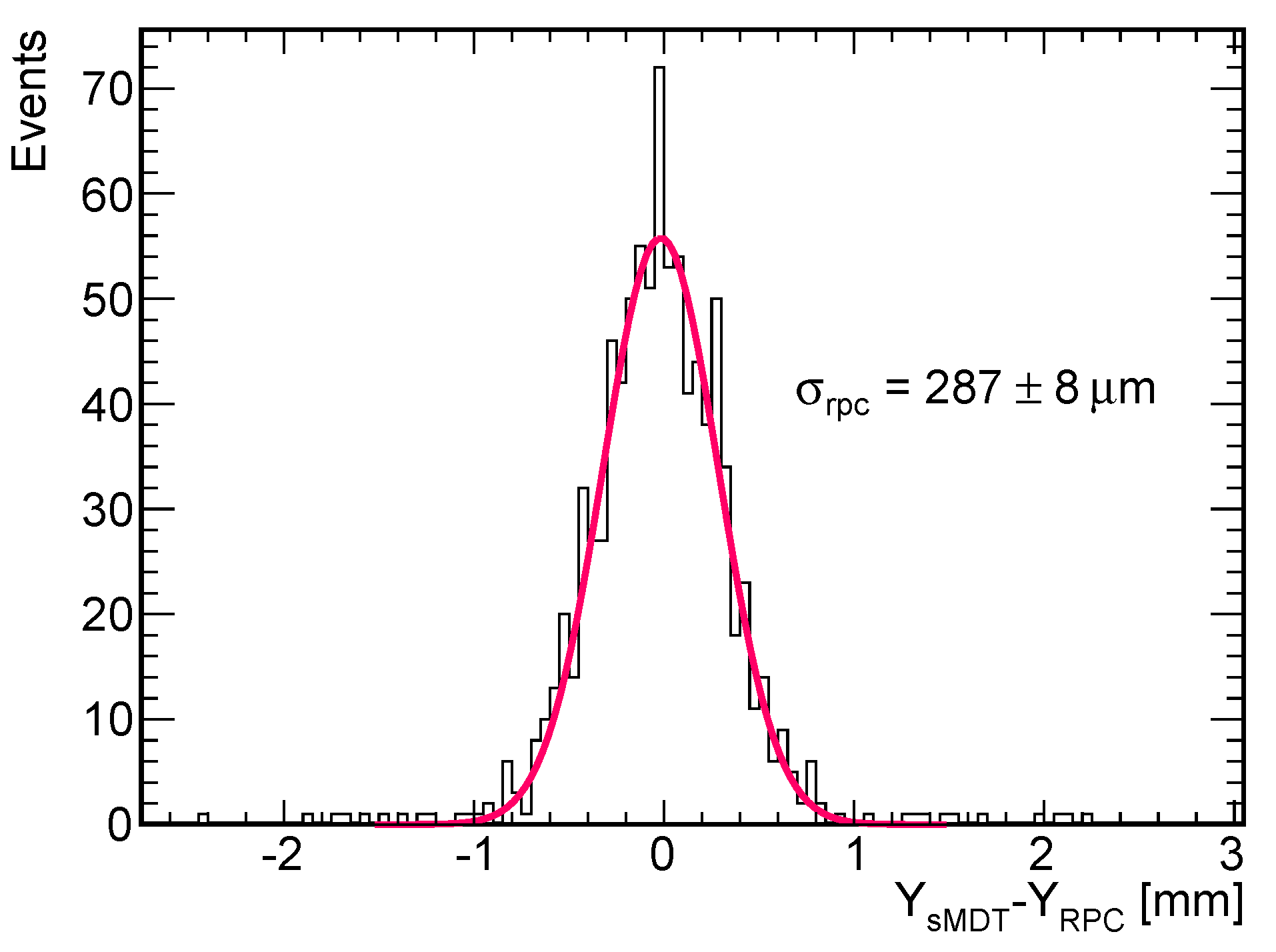}
  \end{center}
  \caption{The difference between RPC reconstructed and predicted
    positions.  The reconstructed positions on RPC only used time
    information and were determined to be the mean positions of the
    fired strip centers. RPC resolution after subtracting 100 $\mu$m
    track prediction uncertainty is shown in the plot. }
  \label{online_space_res}
\end{figure}

For the second approach, only the strip  timing information was 
used. The fired strips were selected within the 20 ns time window
after the earliest arrival time. The hit position was then calculated
as the mean value of the central positions of all selected strips. The
typical distribution of residuals between the predicted positions from
sMDT and the calculated position on the RPC is shown in
Fig~\ref{online_space_res}. The spatial resolution was determined to be 287
$\mu$m after the subtraction of 100 $\mu$m estimated uncertainty on
the predicted position from the sMDT.  
The ability to measure muon hit position with a precision of $\sim 300 \mu$m 
within $10-20$ ns with a read out of only a few strips makes
these RPC detectors excellent high-precision trigger devices for muon detection.

\section{Time resolution and second coordinate measurements}
\label{sec-secondary-coord}
\subsection{Experimental setup and readout electronics}

\begin{figure}[htbp]
  \begin{center}
    \includegraphics[width=1\linewidth]{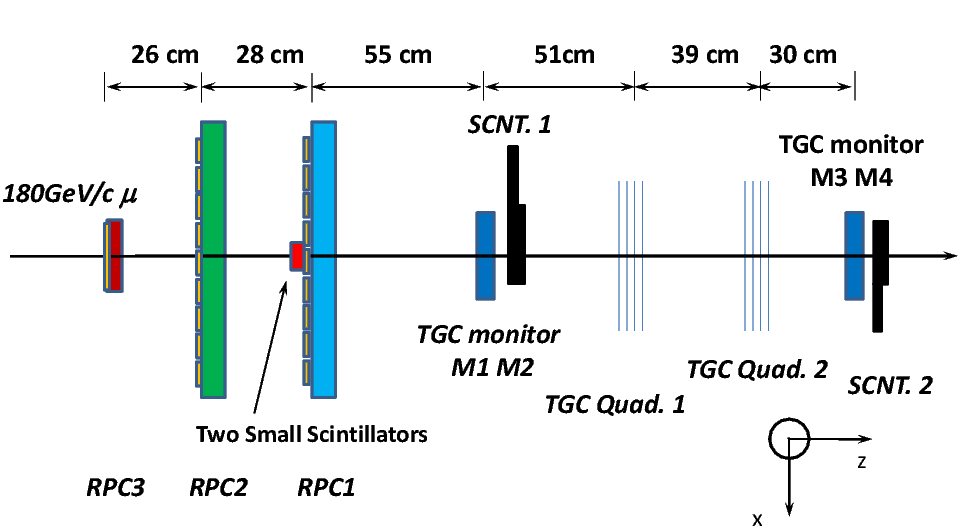}
  \end{center}
  \caption{Side view of the beam line set up for time resolution and
    second coordinate measurements.}
  \label{2012tb_setup}
\end{figure}

The narrow gas-gap RPC time resolution and second coordinate
measurement capability were investigated in the same CERN beam line in
June 2012 with a different experimental setup and readout electronics.
The experimental setup using two glass chambers (denoted as RPC1 and
RPC2) and a Bakelite chamber (denoted as RPC3) is illustrated in
Fig.~\ref{2012tb_setup}. The strips for the two glass chambers were
placed perpendicularly to the strips of the Bakelite chamber.
This layout enabled  use of the Bakelite
chamber to provide reference track hit positions for the glass
chambers along their strip direction. Thin Gap Chamber (TGC)~\cite{tgc} doublets
(2 layers) and quadruplets (4 layers) were placed at the downstream
side of the beam line.  Two big scintillators, 1.2 m apart, provided the global 
trigger for the RPC and TGC DAQ systems.

All three RPCs were read out using the electronics chain composed of
NINO~\cite{Anghinolfi_2004} front-end cards developed for the ALICE
multi-gap RPC time-of-flight detector and TDC modules with a time
resolution of 100 ps.  Each NINO card, containing three eight-channel
NINO chips for amplification, shaping and discriminating, accepts
differential signals from the chamber.  The strips of the two glass
chambers were terminated to the ground through 1 M$\Omega$ resistors,
and both ends were capacitively coupled to the NINO inputs through 1
nF capacitors. For the Bakelite chamber, the strips were read out only
from one end, also through 1 M$\Omega$ decoupling resistors. Since
ground sides of the chambers were equipped with signal pickup boards
and negative high voltage were applied, only signals with negative
polarities were sent to the NINO cards. All positive inputs to the
NINO were AC coupled to the common ground of the chamber.

\subsection{Time resolution} 

\begin{figure}[htbp]
  \begin{center}
    \includegraphics[width=0.9\linewidth]{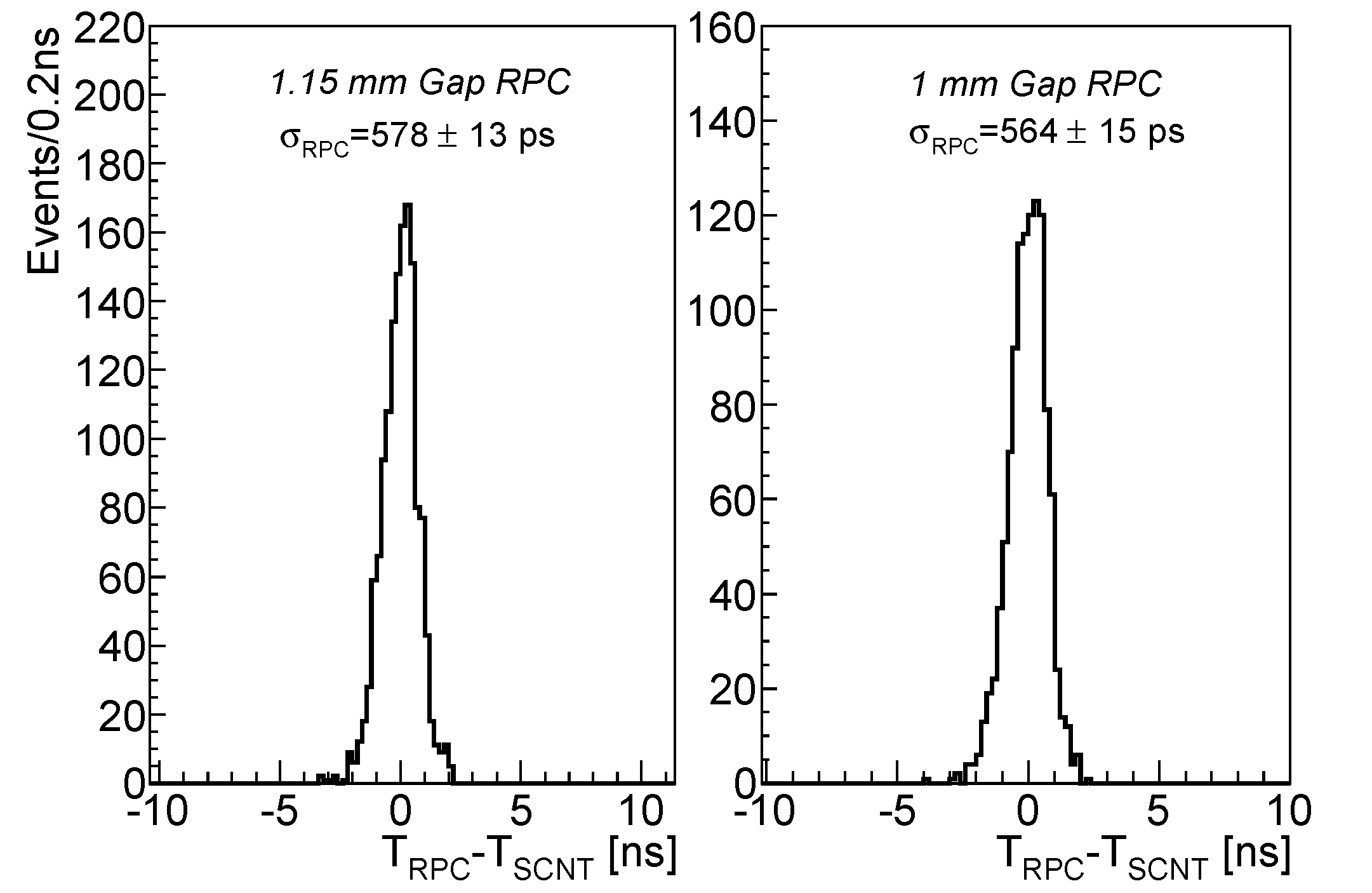}
  \end{center}
  \caption{Time distribution for 1.15 mm gap glass RPC operated at 6.8
    kV (left) and 1 mm gap Bakelite RPC operated at 6.2 kV
    (right). RPC time resolutions after subtracting the scintillator
    time jitters are shown.}
  \label{time_res}
\end{figure}

Time resolutions for 1.15 mm gas-gap glass chambers and 1 mm gas-gap
Bakelite chamber were measured using the average time of the two small
scintillators as the reference time. Since the time jitter for a
single scintillator was measured to be 584 $\pm$ 6 ps, the meantime
of the two scintillators has an uncertainty of 413 $\pm$ 4 ps for the
overall measured time jitter. Time resolutions of 578 $\pm$ 13 ps for
the 1.15 mm gas-gap glass chamber and 564 $\pm$ 15 ps for the 1 mm
gas-gap Bakelite chamber were derived after removing the contribution
from the reference time provided by the two
scintillators. Distributions of the scintillator-corrected RPC time
resolution are shown in Fig.~\ref{time_res}. The charge information
was used later to perform time-walk corrections. Such corrections
yield a time resolution of 510 $\pm$ 10 ps for 1.15 mm gas-gap glass
chamber and 453 $\pm$ 12 ps for the 1 mm gas-gap Bakelite chamber.

\subsection{Meantime measurement} 

The mean value (meantime) of the signal arrival times from both ends
of each readout strip was measured for the glass chambers at different
positions along the strips. An exact location of the hit position
along the glass chamber strip was measured independently using the Bakelite chamber.

\begin{figure}[htbp]
  \begin{center}
    \includegraphics[width=0.9\linewidth]{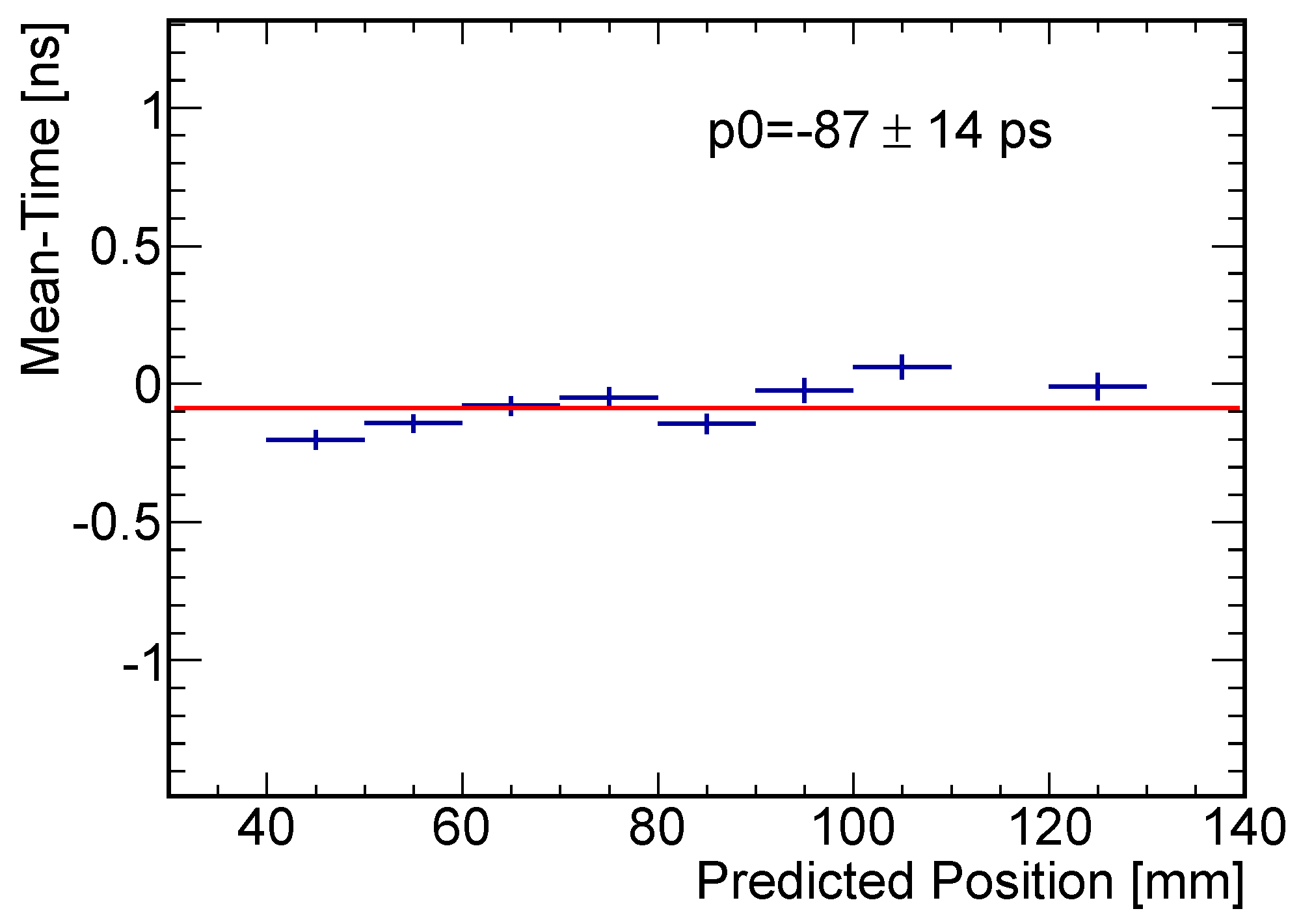}
  \end{center}
  \caption{Meantime of a typical strip as a function of the muon hit
    position along the strip.  The meantime is independent of the hit
    position as expected. The vertical scale has an arbitrary offset.}
  \label{meantime}
\end{figure}

The meantime as a function of the hit position is shown in
Fig.~\ref{meantime} and shows no clear dependence between the
meantime and the hit position. The mean value of the meantime
distribution varies from strip to strip as expected due to different
delays from the readout cables and circuits. After compensating for
these delays and offsets channel-by-channel through a programmable
logic circuit, it is possible to latch the coincidence window between
multiple RPC layers within a few nanoseconds.  With these timing
corrections the trigger coincidence window can be minimized to reduce
background in a high-rate environment.

\subsection{Time difference and second coordinate measurements}

\begin{figure}[htbp]
  \begin{center}
    \includegraphics[width=1\linewidth]{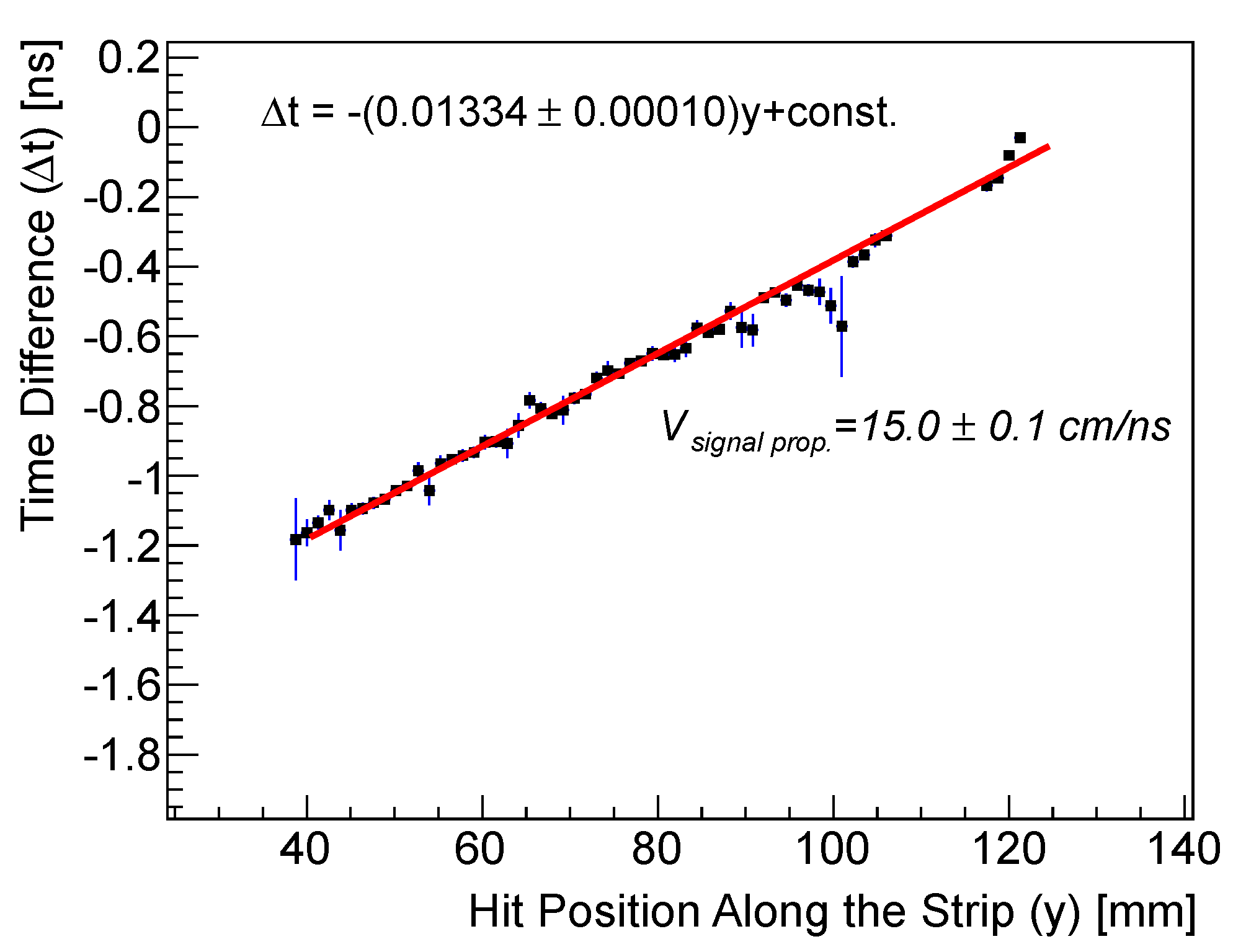}
  \end{center}
  \caption{The time difference calculated using the arrival time
    information from both ends of the readout strips as a function of
    the predicted muon hit position.}
  \label{sig_vel}
\end{figure}

Time differences ($\Delta$t) from the two ends of the readout strips
were measured, and the dependence on the muon hit position along the
strip ($y$) is shown in Fig.~\ref{sig_vel}. A linear fit of this
dependence yields an estimation of the signal propagation speed of
15.0 cm/ns along the strip. Slight differences of the measured
transmission speed, within 1.5 cm/ns, were observed for different
channels. The resolution of the time difference distribution for a
typical strip was measured to be 150 ps which was close to the 100 ps
time resolution of the TDC module used.

\begin{figure}[htbp]
  \begin{center}
    \includegraphics[width=1\linewidth]{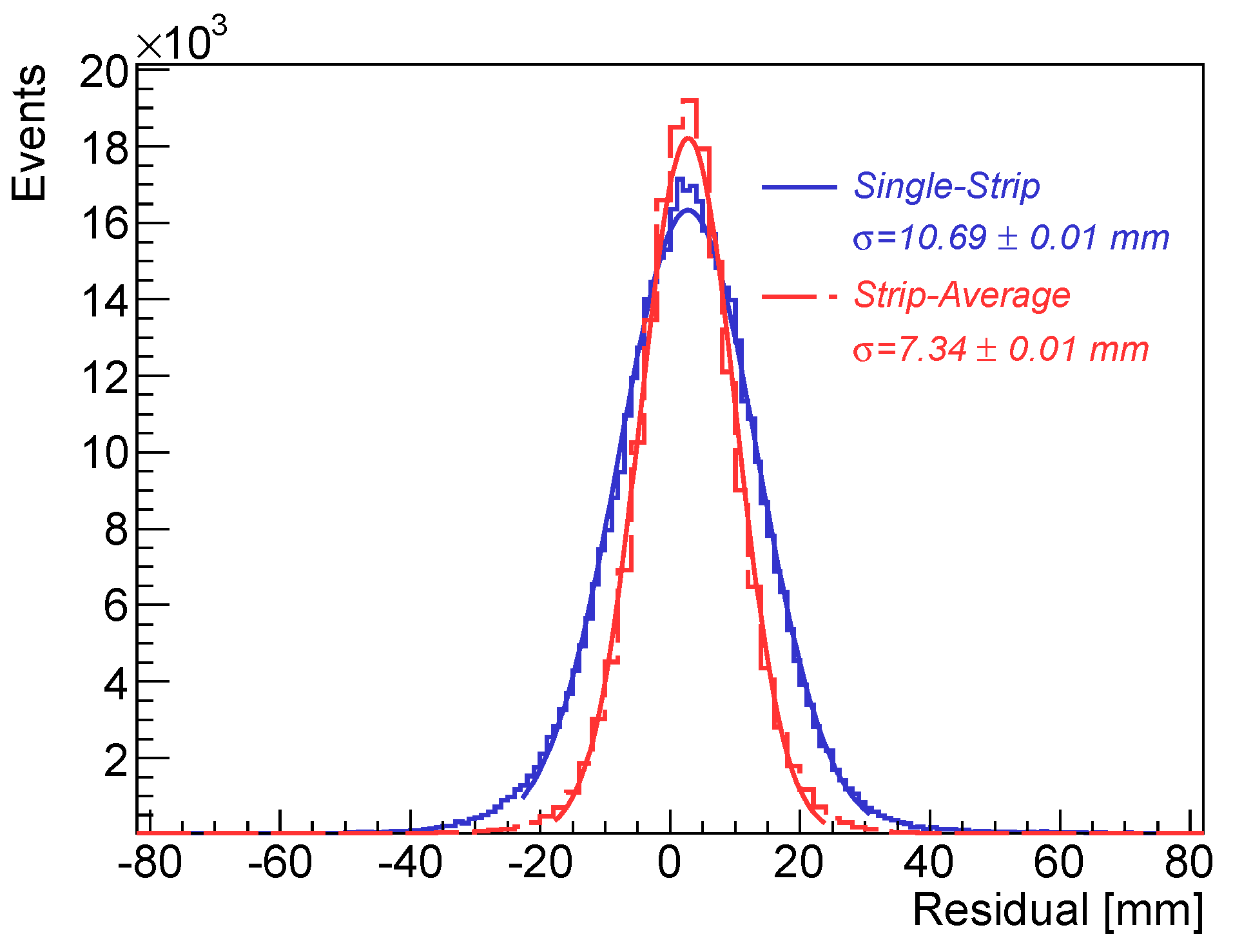}
  \end{center}
  \caption{Spatial resolution along the strip using the time
    difference information from both ends of the readout strips. Hit
    position is determined using the time difference measured in a
    single strip (blue dashed histogram) and the average time
    difference from multiple strips (red bashed histogram). These two
    histograms are fitted using Gaussian functions.}
  \label{space_res_tdiff}
\end{figure}

The spatial resolution along the strip direction in the glass RPC was
measured by comparing the reconstructed positions in the glass RPC
with the positions measured using the Bakelite RPC. Since the Bakelite chamber
was also equipped with 1.27 mm-pitch readout strips, the predicted muon
hit positions have sub-mm precision. For the glass RPC, the hit
position was calculated as the product of the time difference from
both ends and the signal propagation velocity divided by two. Two
approaches were used for the determination of the muon hit
position. One method only used the time difference information from a
single strip.  Another approach determined the hit position by
averaging the values from multiple strips within the
cluster. The residuals between the predicted and the reconstructed
positions are shown in Fig.~\ref{space_res_tdiff} for both
approaches. Resolutions along the strip direction are found to be
10.69 mm using a single strip and 7.34 mm using multiple strips.

\section{Conclusions}
\label{sec-conclusion}

We have presented new developments of fast and high-precision trigger
and tracking using RPCs.  Two beam tests were
performed with 1.0 mm gap Bakelite electrode and 1.15 mm gap glass
electrode RPCs equipped with 1.27 mm fine-pitch strips and read out
from both ends of the strips using charge ADCs and fast TDCs. 
 The RPC time resolution was measured to be better than 600 ps.  
Spatial resolutions for the primary coordinate were experimentally
measured to be better than 200 $\mu$m using charge information and 287
$\mu$m using hit arrival time information.  Signal arrival time
differences from both ends of the readout strips were used for the
determination of second coordinates of incident  muons, and resolutions
better than 7.5 mm was achieved using TDCs with a time resolution of 100 ps.

The fine segmentation in the two-dimensional space, along with the
excellent timing capability and uniform meantime for the signal from both
ends make narrow-gap, fine-pitch RPCs an attractive high-precision trigger device 
for future lepton and hadron colliders.

\section{Acknowledgments}

We would like to thank M. Lippert and P. Schwegler from the
Max Plank Institute, and G. Mikenberg, M. Shoa and their colleagues
from the ATLAS TGC group for their help during the beam tests. The
authors would also like to acknowledge M.C.S. Williams and R. Zouevski
for their help using NINO front-end electronics.  This work is supported 
in part by the Department of Energy under contracts DE-SC0007859 and DE-AC02-98CH10886, 
and by National Science Foundation of China under contract 11025528.

\end{document}